\documentclass[12pt]{article}
\pdfoutput=1

\usepackage{amsmath}
\usepackage{amssymb}
\usepackage{epsfig}
\usepackage{epstopdf}
\usepackage{latexsym}
\usepackage{color}
\numberwithin{equation}{section}
\usepackage{mathrsfs} 

\usepackage[
      colorlinks=false,
      linkcolor=darkblue,  
      urlcolor=blue,    
      filecolor=blue,     
      citecolor=red,
linktocpage=true,
      pdfstartview=FitV,
      bookmarksopen=true    
      ]{hyperref}

\begin{document}

\begin{titlepage}

\centerline
\centerline
\centerline
\bigskip
\bigskip
\centerline{\Huge \rm $AdS$ solutions with spindle factors}
\bigskip
\bigskip
\bigskip
\bigskip
\bigskip
\bigskip
\bigskip
\bigskip
\bigskip
\centerline{\rm Minwoo Suh}
\bigskip
\centerline{\it School of General Education, Kumoh National Institute of Technology,}
\centerline{\it Gumi, 39177, Korea}
\bigskip
\centerline{\tt minwoosuh1@gmail.com} 
\bigskip
\bigskip
\bigskip
\bigskip
\bigskip
\bigskip
\bigskip
\bigskip

\begin{abstract}
\noindent We provide the spindle interpretations of previously known solutions, $AdS_3\times\Sigma\times{S}^1\times{KE}_2^+\times{T}^2$ of type IIB supergravity and $AdS_2\times\Sigma\times{S}^1\times{KE}_4^+\times{T}^2$ of eleven-dimensional supergravity, where $\Sigma$ is a spindle factor. For the $AdS_3$ solutions, the internal space of $\Sigma\times{S}^1\times{KE}_2^+$ was previously known as the $\mathscr{Y}^{p,q}$ manifold, which shares the identical topology of the $Y^{p,q}$ manifolds. Unlike the previously known spindle solutions, the gauged supergravity origin and the field theory dual of the solutions are not clear at the moment. We perform the flux quantizations and calculate the holographic central charge and the Bekenstein-Hawking entropy of the presumed black hole solutions, respectively.
\end{abstract}

\vskip 6cm

\flushleft {September, 2024}

\end{titlepage}

\tableofcontents

\section{Introduction}

Recently, there has been development in constructing and understanding the geometry of anti-de Sitter solutions obtained by wrapping branes on an orbifold. A particular example of such orbifolds is a spindle which is a weighted projective space, $\mathbb{WCP}_{[n_-,n_+]}^1$, with orbifold singularities at two poles. Especially, the ``minimal" spindle solutions, \cite{Ferrero:2020laf, Ferrero:2020twa}, from four- and five-dimensional minimal gauged supergravity were obtained by introducing a new global completion of previously known solutions of Gauntlett-Kim (GK) geometry. These are classes of supersymmetric $AdS_3\times{Y}_7$ solutions of type IIB supergravity, \cite{Kim:2005ez}, and $AdS_2\times{Y}_9$ solutions of eleven-dimensional supergravity, \cite{Kim:2006qu}, with an Abelian R-symmetry. The odd dimensional manifolds of $Y_{2n+1}$, $n\ge3$, are referred to the GK geometries, \cite{Gauntlett:2007ts}. Some explicit solutions of the GK geometry were constructed in \cite{Gauntlett:2006af, Gauntlett:2006ns, Donos:2008ug, Donos:2008hd} and the local forms of the minimal spindle solutions were previously given in \cite{Gauntlett:2006af, Gauntlett:2006ns}.

In this paper, we provide the spindle interpretations of previously known solutions. To be specific, we consider the $AdS_3\times{Y}_7$ solutions of type IIB supergravity and $AdS_2\times{Y}_9$ solutions of eleven-dimensional supergravity obtained in section 4 and 5 of \cite{Donos:2008ug}, respectively. These solutions, in a sense, generalize the solutions of \cite{Gauntlett:2006af, Gauntlett:2006ns} by extra fluxes from turning on Chern-Simons or ``transgression" terms in the action. Overall, compared to the topologies of solutions in \cite{Gauntlett:2006af, Gauntlett:2006ns},
\begin{align}
AdS_3\times\Sigma\times{S}^1\times{KE}_4^+\,,\\ \notag
AdS_2\times\Sigma\times{S}^1\times{KE}_6^+\,,
\end{align}
the solutions of \cite{Donos:2008ug} are 
\begin{align}
AdS_3\times\Sigma\times{S}^1\times{KE}_2^+\times{T}^2\,,\\ \notag
AdS_2\times\Sigma\times{S}^1\times{KE}_4^+\times{T}^2\,,
\end{align}
where $KE_d^+$ are $d$-dimensional K\"ahler-Einstein manifolds with positive curvature, $T^2$ is a two-torus, and $\Sigma$ is the spindle. For the $AdS_3$ solutions, the global completion of the local solution and the calculation of holographic central charge was already performed in \cite{Donos:2008ug}. We reinterpret the analysis to reveal the structure of the solution with a spindle factor. On the other hand, for the $AdS_2$ solutions, only the local form of the solution was given in \cite{Donos:2008ug}. We perform the global analysis of the solutions and calculate the Bekenstein-Hawking entropy of the presumed black holes having the $AdS_2$ solution as a horizon.

There are two classes of solutions from \cite{Donos:2008ug}: the cases with parameter, $Q\neq0$ and $Q=0$. For the $AdS_3$ solutions, when $Q=0$, the three-form flux vanishes identically and only the self-dual five-form flux is non-trivial. We find spindle solutions for both $Q\neq0$ and $Q=0$ cases, but, for the flux quantization and the calculations of observables, we will only consider the $Q=0$ case. For the $AdS_3$ solutions in the non-spindle form with $Q\neq0$ in \cite{Donos:2008ug} , the flux quantization and the calculation of holographic central charge was carefully performed in \cite{Donos:2008hd}.

However, unlike the previously known spindle solutions, the gauged supergravity origin of the spindle solutions from \cite{Donos:2008ug} is not clear at the moment. As mentioned, the minimal spindle solutions, $AdS_3\times\Sigma$ and $AdS_2\times\Sigma$, from \cite{Gauntlett:2006af, Gauntlett:2006ns} are solutions of four- and five-dimensional minimal gauged supergravity, respectively, \cite{Ferrero:2020laf, Ferrero:2020twa}. Hence, they are naturally proposed to be the near horizon geometries of D3- and M2-branes wrapped on a spindle and holographically dual to 4d $\mathcal{N}$=4 SYM theory and 3d ABJM theory compactified on a spindle, respectively. On the other hand, the field theory duals of the spindle solutions from \cite{Donos:2008ug} are not clear at this point.{\footnote {A part of the internal space was named, $\mathscr{Y}^{p,q}=\Sigma\times{S}^1\times{KE}_2^+$. As $\mathscr{Y}^{p,q}$ has the same topology of the five-dimensional Sasaki-Einstein manifold, $Y^{p,q}$, the field theory dual of $AdS_3\times{T}^2\times\mathscr{Y}^{p,q}$ solutions of \cite{Donos:2008ug} was proposed to be 4d $Y^{p,q}$ quiver gauge theories, \cite{Benvenuti:2004dy}, compactified on a two-torus, $T^2$, \cite{Benini:2015bwz}. Furthermore, the holographic central charge was shown to agree with the central charge from the field theory calculations, \cite{Benini:2015bwz}. However, this proposal was disputed in \cite{Couzens:2017nnr, Couzens:2018wnk}, as it is $p\ge{q}$ for $Y^{p,q}$, but $q>p$ for $\mathscr{Y}^{p,q}$.}}

Recently, the five-dimensional Sasaki-Einstein manifold, $Y^{p,q}$, was shown to contain a spindle factor, \cite{Ferrero:2024vmz}. As the part of the internal space, $\mathscr{Y}^{p,q}=\Sigma\times{S}^1\times{KE}_2^+$, has the same topology of $Y^{p,q}$, our result may support the observation made in \cite{Ferrero:2024vmz}.

In section \ref{ads3}, we provide the spindle interpretation of $AdS_3\times\Sigma\times{S}^1\times{KE}_2^+\times{T}^2$ solutions of type IIB supergravity where $\Sigma$ is a spindle factor. For the $Q=0$ case, we perform the flux quantization and calculate the holographic central charge. In section \ref{ads2}, we provide the spindle interpretation of $AdS_2\times\Sigma\times{S}^1\times{KE}_4^+\times{T}^2$ solutions of eleven-dimensional supergravity where $\Sigma$ is a spindle factor. For the $Q=0$ case, we perform the flux quantization and calculate the Bekenstein-Hawking entropy of the presumed black hole solution. We conclude in section \ref{conc}.

\section{$AdS_3$ solutions of type IIB supergravity} \label{ads3}


By completing the square of $D\psi$, we rewrite the metric of the solution of type IIB supergravity in section 4 of \cite{Donos:2008ug} by
\begin{align} \label{tenmet}
\frac{1}{L^2}ds_{10}^2\,=\,\frac{\beta}{y^{1/2}}&\left[ds_{AdS_3}^2+\frac{1}{4\beta^2y^2U}dy^2+\frac{U}{4\left(1-Q^2y^2\right)}dz^2\right. \notag \\
&+\frac{1-Q^2y^2}{4\beta^2}\left(D\psi-\frac{y\left(1-Q^2y\right)}{1-Q^2y^2}dz\right)^2+\left.\frac{1}{\beta^2}ds_{S^2}^2+\frac{y}{\beta^2}ds_{T^2}^2\right]\,,
\end{align}
where we define
\begin{equation} \label{defU}
U(y)\,=\,1-\frac{1}{\beta^2}\left(1-y\right)^2-Q^2y^2\,,
\end{equation}
with
\begin{equation}
D\psi\,=\,d\psi+2V\,, \qquad dV\,=\,2J_{S^2}\,,
\end{equation}
where $ds_{S^2}^2$ is a metric of unit two-sphere and $ds_{T^2}^2$ is a metric of two-torus. The two-sphere, $S^2$, has been chosen as a particular example of two-dimensional K\"ahler manifold with positive curvature, $KE_2^+$. The two-sphere is normalized to have $R_{S^2}=4J_{S^2}$ where $R_{S^2}$ and $J_{S^2}$ are the Ricci and the K\"ahler forms on $S^2$, respectively. $L$ is an arbitrary length scale and $\beta$ and $Q$ are constants. We have $\psi\in[0,2\pi]$ and the ranges of $y$ and $z$ will be determined in the following. The self-dual five-form flux is given by
\begin{align} \label{tenflux}
F_{(5)}\,=\,\text{vol}_{AdS_3}\wedge\omega_2+\omega_5\,,
\end{align}
where we have
\begin{equation}
\frac{1}{L^4}\omega_2\,=\,\frac{\beta}{2y^2}dy\wedge{d}z+2J_{S^2}+2\text{vol}_{T^2}\,,
\end{equation}
and
\begin{align} 
\frac{1}{L^4}\omega_5\,=\,-&\frac{y\left(1-Q^2y\right)}{\beta^2}\text{vol}_{T^2}\wedge{J}_{S^2}\wedge{d}z+\frac{1-Q^2y^2}{\beta^2}\text{vol}_{T^2}\wedge{J}_{S^2}\wedge{D}\psi \notag \\
-&\frac{1}{4\beta^2y^2}dy\wedge{D}\psi\wedge{J}_{S^2}\wedge{d}z-\frac{1}{4\beta^2}\text{vol}_{T^2}\wedge{d}y\wedge{D}\psi\wedge{d}z\,,
\end{align}
and $\text{vol}_{S^2}$ and $\text{vol}_{T^2}$ are the volume forms on $S^2$ and $T^2$, respectively. The three-form flux is given by
\begin{equation}
\frac{1}{L^2}G_{(3)}\,=\,\frac{Q}{\beta}d\bar{u}\wedge\left[\frac{1}{2}dy\wedge{D}\psi-\frac{1}{2}dy\wedge{d}z+2yJ_{S^2}\right]\,,
\end{equation}
where a complex coordinate, $u=u^1+iu^2$, is introduced on two-torus with $ds^2_{T^2}=idud\bar{u}$. There are largely two classes of solutions: $Q\neq0$ and $Q=0$. When we have $Q=0$, the three-form flux vanishes and only the five-form flux is non-trivial. Later, for the flux quantizations and the calculation of holographic central charge, we will only consider the case with $Q=0$.

Now, from the metric in \eqref{tenmet}, we consider the two-dimensional surface,
\begin{equation}
ds_\Sigma^2\,=\,\frac{1}{4\beta^2y^2U}dy^2+\frac{U}{4\left(1-Q^2y^2\right)}dz^2\,,
\end{equation}
with the gauge field on the surface which fibers $D\psi$,
\begin{equation}
A\,=\,-\frac{y\left(1-Q^2y\right)}{1-Q^2y^2}dz\,.
\end{equation}
Unlike the minimal spindle solution in \cite{Ferrero:2020laf}, $AdS_3\times\Sigma$ is not a solution of five-dimensional minimal gauged supergravity. There are two roots, $0<y_1<y_2$, of $U(y)=0$,
\begin{equation}
y_1\,=\,\frac{1-\beta\sqrt{1+Q^2\left(\beta^2-1\right)}}{1+Q^2\beta^2}\,, \qquad y_2\,=\,\frac{1+\beta\sqrt{1+Q^2\left(\beta^2-1\right)}}{1+Q^2\beta^2}\,,
\end{equation}
where the parameters are restrained by
\begin{equation}
0<\beta^2<1\,, \qquad 0\le{Q}^2\le\frac{1}{1-\beta^2}\,.
\end{equation}
We take $y\in[y_1,y_2]$ to have a positive definite metric on $\Sigma$.

Approaching the roots of $y_*\,=y_{1,2}$, the metric becomes
\begin{equation}
ds_\Sigma^2\,\approx\,\frac{1}{-\beta^2y_*^2U'(y_*)}\left[dR^2+\frac{\beta^2y_*^2U'(y_*)^2}{4\left(1-Q^2y_*^2\right)}R^2dz^2\right]\,.
\end{equation}
The metric becomes $\mathbb{R}^2/\mathbb{Z}_n$, if the coordinate, $z$, has period $\Delta{z}$ with
\begin{equation} \label{npnm}
\frac{\beta{y}_1U'(y_1)}{2\sqrt{1-Q^2y_1^2}}\,=\,\frac{2\pi}{\Delta{z}\,n_+}\,, \qquad \frac{\beta{y}_2U'(y_2)}{2\sqrt{1-Q^2y_2^2}}\,=\,-\frac{2\pi}{\Delta{z}\,n_-}\,,
\end{equation}
where $n_\pm$ are coprime positive integers. The Euler characteristic of the surface is given by
\begin{align}
\chi(\Sigma)\,=&\,\frac{1}{4\pi}\int_\Sigma{R}_\Sigma\text{vol}_\Sigma \notag \\
=&\,\left.-\frac{\beta{y}\left(1-Q^2y^2\right)U'(y)+2\beta{Q}^2y^2U(y)}{\left(1-Q^2y^2\right)^{3/2}}\frac{\Delta{z}}{4\pi}\right|^{y_2}_{y_1} \notag \\
=&\,\frac{\Delta{z}}{4\pi}\left(-\frac{\beta{y}U'(y_2)}{\sqrt{1-Q^2y_2^2}}+\frac{\beta{y}U'(y_1)}{\sqrt{1-Q^2y_1^2}}\right) \notag \\
=&\,\frac{1}{n_+}+\frac{1}{n_-}\,.
\end{align}
The charge quantization gives
\begin{align} \label{chargeq}
\mathcal{Q}\,=&\,\frac{1}{2\pi}\int_\Sigma{dA} \notag \\
=&\,\frac{1}{2\pi}\int_\Sigma{d}\left(-\frac{y\left(1-Q^2y\right)}{1-Q^2y^2}\right)\wedge{dz} \notag \\
=&\,-\frac{\Delta{z}}{\pi}\frac{\beta\sqrt{1-Q^2\left(1-\beta^2\right)}}{1-Q^2} \notag \\
=&\,\frac{\Delta{z}}{4\pi}\left(\frac{\beta{y}U'(y_2)}{\sqrt{1-Q^2y_2^2}}+\frac{\beta{y}U'(y_1)}{\sqrt{1-Q^2y_1^2}}\right) \notag \\
=&\,\frac{1}{n_+}-\frac{1}{n_-}\,,
\end{align}
where, from the third to the fourth line, due to the complexity of expressions, we only checked the equivalence by plugging in some numerical values. These characterize the two-dimensional surface to be topologically a two-sphere with two conical singularities at the north and south poles, which is known as the spindle. In particular, from \eqref{chargeq}, the spindle solutions are in the anti-twist class, \cite{Ferrero:2021etw}.

By solving the equations in \eqref{defU} and \eqref{npnm}, 
\begin{align}
-\left.\frac{\beta{y}_1U'(y_1)}{2\sqrt{1-Q^2y_1^2}}\right/\frac{\beta{y}_2U'(y_2)}{2\sqrt{1-Q^2y_2^2}}\,=&\,\frac{n_-}{n_+}\,, \notag \\
U(y_1)\,=\,1-\frac{1}{\beta^2}\left(1-y_1\right)^2-Q^2y_1^2\,=&\,0\,, \notag \\
U(y_2)\,=\,1-\frac{1}{\beta^2}\left(1-y_2\right)^2-Q^2y_2^2\,=&\,0\,,
\end{align}
we express the roots and the parameter in terms of $n_+$, $n_-$, and $Q$,
\begin{align}
y_1\,=&\,\frac{2n_-}{n_++n_--\left(n_+-n_-\right)Q^2}\,, \notag \\
y_2\,=&\,\frac{2n_+}{n_++n_-+\left(n_+-n_-\right)Q^2}\,, \notag \\
\beta\,=&\,\frac{\left(n_+-n_-\right)\sqrt{1-Q^2}}{\sqrt{\left(n_++n_-\right)^2-\left(n_+-n_-\right)^2Q^2}}\,.
\end{align}
We determine the period of $z$ by plugging above into \eqref{npnm},
\begin{equation}
\Delta{z}\,=\,\pi\frac{\left(n_++n_-\right)^2-\left(n_+-n_-\right)^2Q^2}{n_+n_-\left(n_++n_-\right)}\,.
\end{equation}

For the flux quantization and the calculation of holographic central charge in the following, we consider the case of
\begin{equation}
Q\,=\,0\,,
\end{equation}
which turns off the three-from flux. 

Now we consider the flux quantization of the five-form flux,
\begin{equation}
N(D)\,=\,\frac{1}{\left(2\pi\ell_s\right)^4g_s}\int_DF_{(5)}\,\in\,\mathbb{Z}\,,
\end{equation}
for any five-cycle, $D\in{H}_5(Y_7,\mathbb{Z})$, in $AdS_3\times{Y}_7$. $\ell_s$ and $g_s$ are the string length and the string coupling constant, respectively. We introduce two integers, $N$ and $M$, by{\footnote{We denote volume forms by $\text{vol}$ and volumes by $\it{vol}$.}}
\begin{align}
\frac{L^4}{64\pi\ell_s^4g_s}\,=&\,\frac{n_+^2n_-^2\left(n_+-n_-\right)}{\left(n_++n_-\right)^4}N\,, \notag \\
vol_{T^2}\,=&\,\frac{\pi}{4}\frac{\left(n_+-n_-\right)\left(n_++n_-\right)^2}{n_+n_-}\frac{M}{N}\,.
\end{align}
Then the fluxes through five-cycles give
\begin{align}
\frac{1}{\left(2\pi\ell_s\right)^4g_s}\int_{M_5}F_{(5)}\,=&\,-N\,, \notag \\
\frac{1}{\left(2\pi\ell_s\right)^4g_s}\int_{E_1\times{T}^2}F_{(5)}\,=&\,-2n_-M\,, \notag \\
\frac{1}{\left(2\pi\ell_s\right)^4g_s}\int_{E_2\times{T}^2}F_{(5)}\,=&\,-2n_+M\,, \notag \\
\frac{1}{\left(2\pi\ell_s\right)^4g_s}\int_{E_3\times{T}^2}F_{(5)}\,=&\,-\left(n_+-n_-\right)M\,,
\end{align}
where we have $M_5=(y,z,\psi,S^2)$, $E_1=(z,S^2)$ at $y=y_1$, $E_2=(z,S^2)$ at $y=y_2$, and $E_3=(y,z,\psi)$, respectively. Notice that they satisfy the homology relation,
\begin{equation}
E_1=E_2-2E_3\,,
\end{equation}
and see section 4.1 of \cite{Donos:2008ug} for details.

Now we calculate the holographic central charge of dual 2d superconformal field theories. For the metric of the form,
\begin{equation}
ds_{10}^2\,=\,e^{2\mathcal{A}}ds_{AdS_3}^2+ds^2_{M_7}\,,
\end{equation}
the holographic central charge is
\begin{equation}
c\,=\,\frac{3}{2G_N^{(10)}}\int_{M_7}e^\mathcal{A}\text{vol}_{M_7}\,,
\end{equation}
where the ten-dimensional Newton's gravitational constant is $G_N^{(10)}\,=\,2^3\pi^6\ell_s^8$. We obtain
\begin{align}
c\,=&\,\frac{3}{16\pi^6\ell_s^8}\int\frac{1}{8\beta^2y^2}\,dy\wedge{d}z\wedge{D}\psi\wedge{\text{vol}}_{S^2}\wedge{\text{vol}}_{T^2} \notag \\
=&\,-\frac{3}{16\pi^4\ell_s^8}\frac{\Delta{z}}{\beta^2}\left(\frac{1}{y_2}-\frac{1}{y_1}\right)vol_{T^2}\,,
\end{align}
where $vol_{S^2}=4\pi$. We note that the holographic central charge is given by
\begin{equation}
c\,=\,\frac{24n_+n_-\left(n_+-n_-\right)^2NM}{\left(n_++n_-\right)^2}\,.
\end{equation}

We notice that by reparametrizing $(n_+,n_-)$ by
\begin{equation}
n_+\,=\,\frac{p+2q}{2}\,, \qquad n_-\,=\,\frac{p}{2}\,,
\end{equation}
the calculations in this section translate and match the results presented in section 4.1 of \cite{Donos:2008ug}.

\section{$AdS_2$ solutions of eleven-dimensional supergravity} \label{ads2}


By completing the square of $D\psi$, we rewrite the metric of the solution of eleven-dimensional supergravity in section 5 of \cite{Donos:2008ug} by
\begin{align} \label{elevenmet}
\frac{1}{L^2}ds_{11}^2\,=&\,\frac{1}{6^{4/3}\beta^{2/3}y^{4/3}}\left[ds_{AdS_2}^2+\frac{9\beta}{yU}dy^2+\frac{U}{1-Qy^3}dz^2\right. \notag \\
&\left.+4\beta{y}\left(1-Qy^3\right)\left(D\psi+\frac{1-3y+2Qy^3}{2\left(1-Qy^3\right)}dz\right)^2+36\beta{y}ds_{KE_4^+}^2+36\beta{y}^2ds_{T^2}^2\right]\,,
\end{align}
where we define{\footnote{We fixed a typographical error in $g(y)$ in (5.4) of \cite{Donos:2008ug}: $-4\beta{Q}y^3\rightarrow-4\beta{Q}y^4$. Also we correct $J_{S^2}\rightarrow{J}_{KE}$ below (5.2).}}
\begin{equation} \label{defU2}
U(y)\,=\,1-9\beta{y}\left(1-y\right)^2-Qy^3\,,
\end{equation}
with
\begin{equation}
D\psi\,=\,d\psi+2V\,, \qquad dV\,=\,2J_{KE_4^+}\,,
\end{equation}
where $ds_{KE_4^+}^2$ is a metric of four-dimensional K\"ahler-Einstein manifold with positive curvature and $ds_{T^2}^2$ is a metric of two-torus. $KE_4^+$ is normalized to have $R_{KE_4^+}=6J_{KE_4^+}$ where $R_{KE_4^+}$ and $J_{KE_4^+}$ are the Ricci and the K\"ahler forms on $KE_4^+$, respectively. $L$ is an arbitrary length scale and $\beta$ and $Q$ are constants. We have $\psi\in[0,2\pi]$ and the ranges of $y$ and $z$ will be determined in the following. The four-form flux is given by
\begin{equation}
G_{(4)}\,=\,\text{vol}_{AdS_2}\wedge\omega_2+\omega_4\,,
\end{equation}
where we have
\begin{equation}
\frac{1}{L^3}\omega_2\,=\,-J_{KE_4^+}-\frac{2}{y^3}dy\wedge{d}z+\frac{4g(y)}{y^3}dy\wedge{D}\psi-\frac{1}{2}\text{vol}_{T^2}\,,
\end{equation}
and
\begin{align}
\frac{1}{L^3}\omega_4\,=\,6&\beta^{1/2}Q\left(2J_{KE_4^+}\wedge{J}_{KE_4^+}+\frac{1}{3}\left[\left(1-g\right)D\psi-Dz\right]\wedge{J}_{KE_4^+}\wedge{d}y\right. \notag \\
&\left.-2y^2J_{KE_4^+}\wedge\text{vol}_{T^2}-\frac{y}{3}dy\wedge\left[\left(1-g\right)D\psi-Dz\right]\wedge\text{vol}_{T^2}\right)\,,
\end{align}
with
\begin{equation}
g(y)\,=\,-\frac{2\beta{y}\left(1-3y+2Qy^3\right)}{1-8\beta{y}+12\beta{y}^2-4\beta{Q}y^4}\,, 
\end{equation}
and $\text{vol}_{S^2}$ and $\text{vol}_{T^2}$ are the volume forms on $S^2$ and $T^2$, respectively. There are largely two classes of solutions: $Q\neq0$ and $Q=0$. Later, for the flux quantizations and the calculation of the Bekenstein-Hawking entropy, we will only consider the case with $Q=0$. For the case of $Q=0$, the Hodge dual of the four-form flux is
\begin{align}
\frac{1}{L^6}*G_{(4)}\,=\,&6\beta{d}y\wedge{d}z\wedge{D}\psi\wedge*_{KE_4^+}J_{KE_4^+}\wedge\text{vol}_{T^2}+1728\beta^2D\psi\wedge\text{vol}_{KE_4^+}\wedge\text{vol}_{T^2} \notag \\
&+1728\beta\left(3y-1\right)dz\wedge\text{vol}_{KE_4^+}\wedge\text{vol}_{T^2}+\frac{3\beta}{y^2}dy\wedge{d}z\wedge{D}\psi\wedge\text{vol}_{KE_4^+}\,.
\end{align}

We have several choices for four-dimensional K\"ahler-Einstein manifolds: $KE_4^+=\mathbb{CP}^2$ with the Fano index, $I=3$, $KE_4^+=S^2\times{S}^2$ with $I=2$, and del Pezzo surfaces, $KE_4^+=dP_m$, $3\le{m}\le8$, with $I=1$, \cite{Ferrero:2020laf}. Some of the volumes can be obtained from
\begin{equation} \label{ke4vol}
vol_{\mathbb{CP}^n}\,=\,\frac{\pi^n}{n!}\,, \qquad vol_{S^2\times{S}^2}\,=\,16\pi^2\,.
\end{equation}

Now, from the metric in \eqref{elevenmet}, we consider the two-dimensional surface,
\begin{equation}
ds_\Sigma^2\,=\,\frac{9\beta}{yU}dy^2+\frac{U}{1-Qy^3}dz^2\,,
\end{equation}
with the gauge field on the surface which fibers $D\psi$,
\begin{equation}
A\,=\,\frac{1-3y+2Qy^3}{2\left(1-Qy^3\right)}dz\,.
\end{equation}
Unlike the minimal spindle solution in \cite{Ferrero:2020twa}, $AdS_2\times\Sigma$ is not a solution of four-dimensional minimal gauged supergravity. There are three roots, $y_1$ and $0<y_2<y_3$, of $U(y)=0$, but, as they are unwieldy, we do not reproduce them here. We take $y\in[y_2,y_3]$ to have a positive definite metric on $\Sigma$.

Approaching the roots of $y_*\,=y_{2,3}$, the metric becomes
\begin{equation}
ds_\Sigma^2\,\approx\,\frac{36\beta}{-y_*U'(y_*)}\left[dR^2+\frac{y_*U'(y_*)^2}{36\beta\left(1-Qy_*^3\right)}R^2dz^2\right]\,.
\end{equation}
The metric becomes $\mathbb{R}^2/\mathbb{Z}_n$, if the coordinate, $z$, has period $\Delta{z}$ with
\begin{equation} \label{npnm2}
\frac{y_2^{1/2}U'(y_2)}{6\sqrt{\beta\left(1-Qy_2^3\right)}}\,=\,\frac{2\pi}{\Delta{z}\,n_+}\,, \qquad \frac{y_3^{1/2}U'(y_3)}{6\sqrt{\beta\left(1-Qy_3^3\right)}}\,=\,-\frac{2\pi}{\Delta{z}\,n_-}\,,
\end{equation}
where $n_\pm$ are coprime positive integers. The Euler characteristic of the surface is given by
\begin{align}
\chi(\Sigma)\,=&\,\frac{1}{4\pi}\int_\Sigma{R}_\Sigma\text{vol}_\Sigma \notag \\
=&\,\left.\left(-\frac{1}{3}\sqrt{\frac{y}{\beta\left(1-Qy^3\right)}}U'(y)+\frac{Qy^{5/2}U(y)}{\beta^{1/2}\left(1-Qy^3\right)^{3/2}}\right)\frac{\Delta{z}}{4\pi}\right|^{y_3}_{y_2} \notag \\
=&\,\frac{\Delta{z}}{4\pi}\left(-\frac{y_3^{1/2}U'(y_3)}{3\sqrt{\beta\left(1-Qy_3^2\right)}}+\frac{y_2^{1/2}U'(y_2)}{3\sqrt{\beta\left(1-Qy_2^2\right)}}\right) \notag \\
=&\,\frac{1}{n_+}+\frac{1}{n_-}\,.
\end{align}
The charge quantization gives
\begin{align} \label{chargeq2}
\mathcal{Q}\,=&\,\frac{1}{2\pi}\int_\Sigma{dA} \notag \\
=&\,\frac{1}{2\pi}\int_\Sigma{d}\left(\frac{1-3y+2Qy^3}{2\left(1-Qy^3\right)}\right)\wedge{dz} \notag \\
=&\,\frac{\Delta{z}}{4\pi}\left(-\frac{y_3^{1/2}U'(y_3)}{3\sqrt{\beta\left(1-Qy_3^2\right)}}+\frac{y_2^{1/2}U'(y_2)}{3\sqrt{\beta\left(1-Qy_2^2\right)}}\right) \notag \\
=&\,\frac{1}{n_+}+\frac{1}{n_-}\,,
\end{align}
where, from the second to the third line, due to the complexity of expressions, we only checked the equivalence by plugging in some numerical values. These characterize the two-dimensional surface to be topologically a two-sphere with two conical singularities at the north and south poles, which is known as the spindle. In particular, from \eqref{chargeq2}, the spindle solutions are in the twist class, \cite{Ferrero:2021etw}.

For the flux quantization and the calculation of Bekenstein-Hawking entropy in the following, we consider the case of
\begin{equation}
Q\,=\,0\,.
\end{equation}

By solving the equations in \eqref{defU2} and \eqref{npnm2}, 
\begin{align}
-\left.\frac{y_2^{1/2}U'(y_2)}{6\sqrt{\beta}}\right/\frac{y_3^{1/2}U'(y_3)}{6\sqrt{\beta}}\,=&\,\frac{n_-}{n_+}\,, \notag \\
U(y_1)\,=\,1-\frac{1}{\beta^2}\left(1-y_1\right)^2\,=&\,0\,, \notag \\
U(y_2)\,=\,1-\frac{1}{\beta^2}\left(1-y_2\right)^2\,=&\,0\,, \notag \\
U(y_3)\,=\,1-\frac{1}{\beta^2}\left(1-y_3\right)^2\,=&\,0\,, \notag \\
y_1+y_2+y_3\,=&\,2\,,
\end{align}
we express the roots and the parameter in terms of $n_+$ and $n_-$,
\begin{align}
y_1\,=&\,\frac{\left(n_++n_-\right)^2}{3\left(n_+^2-n_+n_-+n_+^2\right)}\,, \notag \\
y_2\,=&\,\frac{\left(n_+-2n_-\right)^2}{3\left(n_+^2-n_+n_-+n_+^2\right)}\,, \notag \\
y_3\,=&\,\frac{\left(2n_+-n_-\right)^2}{3\left(n_+^2-n_+n_-+n_+^2\right)}\,, \notag \\
\beta\,=&\,\frac{3\left(n_+^2-n_+n_-+n_+^2\right)}{\left(n_++n_-\right)^2\left(n_+-2n_-\right)^2\left(2n_+-n_-\right)^2}\,,
\end{align}
We determine the period of $z$ by plugging above into \eqref{npnm2},
\begin{equation}
\Delta{z}\,=\,-\frac{4\pi\left(n_+^2-n_+n_-+n_+^2\right)}{3n_+n_-\left(n_+-n_-\right)}\,,
\end{equation}
where the parameters are restrained by
\begin{equation}
2n_+-n_->0\,, \qquad n_+-2n_-<0\,, \qquad \Leftrightarrow \qquad \frac{n_-}{2}<n_+<2n_-\,,
\end{equation}
which also imply $n_+>0$ and $n_->0$.

Now we consider the flux quantization of the Hodge dual of four-form flux,
\begin{equation}
N(D)\,=\,\frac{1}{\left(2\pi\ell_p\right)^6}\int_D*G_{(4)}\,\in\,\mathbb{Z}\,,
\end{equation}
for any seven-cycle, $D\in{H}_7(Y_9,\mathbb{Z})$, in $AdS_2\times{Y}_9$ and $\ell_p$ is the Planck length. We introduce two integers, $N$ and $M$, by
\begin{align} \label{MN}
\frac{27L^3}{8\pi^4\ell_p^6}vol_{KE_4^+}\,=&\,\frac{n_+\left(2n_+^2-3n_+n_-+n_-^2\right)\left(n_++n_-\right)^4\left(n_+-2n_-\right)^4}{\left(n_+^2-n_+n_-+n_+^2\right)^5}N\,, \notag \\
vol_{T^2}\,=&\,\frac{9\pi}{2}n_-\left(2n_+-n_-\right)^3\left(n_+^2-n_+n_-+n_+^2\right)\frac{M}{N}\,.
\end{align}
Although there is an ambiguity in defining $N$ and $M$, we have chosen the minimal $M$ which makes the following fluxes to be integers. Then the fluxes through the seven-cycles give
\begin{align}
\frac{1}{\left(2\pi\ell_p\right)^6}\int_{yz\psi{KE}_4^+}G_{(4)}\,=&\,N\,, \notag \\
\frac{1}{\left(2\pi\ell_p\right)^6}\int_{\psi{KE}_4^+T^2}G_{(4)}\,=&\,648n_+n_-\left(n_+-n_-\right)\left(n_+^2-n_+n_-+n_+^2\right)^3M\,, \notag \\
\frac{1}{\left(2\pi\ell_p\right)^6}\int_{yz\psi{J}_{{KE}_4^+}T^2}G_{(4)}\,=&\,n_-\left(n_++n_-\right)^2\left(n_+-2n_-\right)^2\left(2n_+-n_-\right)^3M\frac{\pi\int_{\Sigma_a}J_{KE_4^+}}{vol_{KE_4^+}}\,.
\end{align}
We note that the integral of the K\"ahler form over a two-cylce, $\Sigma_a\in{KE}_4^+$, gives
\begin{equation}
\int_{\Sigma_a}J_{KE_4^+}\,=\,2\pi{I}n_a\,,
\end{equation}
where $I$ is the Fano index of $KE_4^+$ and $n_a$ are coprime integers, $e.g.$, see (4.42) and its explanation in \cite{Ferrero:2020twa}. For some K\"ahler-Einstein manifolds, the volumes are given in \eqref{ke4vol}. We also find
\begin{equation}
\frac{1}{\left(2\pi\ell_p\right)^6}\int_{zKE_4^+T^2}G_{(4)}\,=\,\pm432n_\mp\left(n_+-n_-\right)\left(n_++n_-\right)^2\left(n_+-2n_-\right)^2\left(2n_+-n_-\right)^2M\,,
\end{equation}
at $y=y_2$ and $y=y_3$, respectively. 

Now we calculate the Bekenstein-Hawking entropy of the presumed black hole. For the metric of the form,
\begin{equation}
ds_{11}^2\,=\,e^{2\mathcal{A}}\left(ds_{AdS_2}^2+ds^2_{M_9}\right)\,,
\end{equation}
the Bekenstein-Hawking entropy is
\begin{equation}
S_{\text{BH}}\,=\,\frac{1}{G_N^{(11)}}\int_{M_9}e^{9\mathcal{A}}\text{vol}_{M_9}\,,
\end{equation}
where the eleven-dimensional Newton's gravitational constant is $G_N^{(11)}\,=\,\frac{\left(2\pi\right)^8\ell_p^9}{16\pi}$. We obtain
\begin{equation}
S_{\text{BH}}\,=\,\frac{8\sqrt{6}\pi^2}{\sqrt{vol_{KE_4^+}}}n_-\left(n_++n_-\right)^2\left(n_+-2n_-\right)^2\left(2n_+-n_-\right)^3\sqrt{\frac{n_+\left(2n_+^2-3n_+n_-+n_-^2\right)}{n_+^2-n_+n_-+n_-^2}}MN^{1/2}\,.
\end{equation}
In order to verify our choice of $N$ and $M$ in \eqref{MN}, it would be nice to reproduce the Bekenstein-Hawking entropy from field theory or other gravitational calculations which we discuss in the conclusions.

\section{Conclusions} \label{conc}

In this paper, we provided the spindle interpretations of $AdS_3\times\Sigma\times{S}^1\times{KE}_2^+\times{T}^2$ solutions of type IIB supergravity and $AdS_2\times\Sigma\times{S}^1\times{KE}_4^+\times{T}^2$ solutions of eleven-dimensional supergravity where $\Sigma$ is a spindle factor. For the $Q=0$ cases, we performed the flux quantizations and calculated the holographic central charge and the Bekenstein-Hawking entropy of the presumed black hole solutions, respectively. Unlike the previously known spindle solutions, the gauged supergravity origins of these solutions are unclear at this point.

We noted that the $AdS_3$ solutions are in the class of anti-twist like the minimal spindle solutions from D3-branes, \cite{Ferrero:2020laf}. On the other hand, the $AdS_2$ solutions are in the class of twist, unlike the minimal spindle solutions from M2-branes, \cite{Ferrero:2020twa}. It would be interesting to understand the relations between the solutions and the discrepancy in the twist classes.

As discussed in the introduction, it would be interesting to calculate the central charge and the Bekenstein-Hawking entropy of the spindle solutions with $Q\neq0$. As mentioned, for the $AdS_3$ solutions in a non-spindle setup, it was carefully done in \cite{Donos:2008hd}.

Also, as it was discussed in the introduction, it would be most interesting if we could identify the dual field theories of the solutions. If there is any, identifying the gauged supergravity origins of the spindle solutions would be very instructive for identifying the dual field theories. Along the way, calculating the holographic central charge and the Bekenstein-Hawking entropy of the solutions via the extremization principles, \cite{Couzens:2018wnk, Gauntlett:2019pqg}, the gravitational blocks, \cite{Hosseini:2019iad, Faedo:2021nub}, or the equivariant localization, \cite{BenettiGenolini:2023kxp, BenettiGenolini:2023ndb} and \cite{Martelli:2023oqk, Colombo:2023fhu}, might provide some hints for the dual field theories.

\bigskip
\bigskip
\leftline{\bf Acknowledgements}
\noindent We would like to thank Christopher Couzens, Hyojoong Kim, and Nakwoo Kim for interesting discussions. This research was supported by Kumoh National Institute of Technology (2023~2025).

\bibliographystyle{JHEP}
\bibliography{20211220_ref}

\end{document}